\documentclass{aastex}
\makeatletter
\def\lnyoro{\mathrel{\mathpalette\gl@align<}}
\def\gnyoro{\mathrel{\mathpalette\gl@align>}}
\def\gl@align#1#2{\lower.6ex\vbox{\baselineskip\z@skip\lineskip\z@\ialign{$\m@th
#1\hfil##\hfil$\crcr#2\crcr\sim\crcr}}}
\makeatother

\begin{document}

\title{\bf COLOR GRADIENTS IN EARLY-TYPE GALAXIES IN CLUSTERS AT THE
REDSHIFT FROM 0.37 TO 0.56}

\author{Naoyuki Tamura and Kouji Ohta}
\affil{Department of Astronomy, Faculty of Science, Kyoto University,
Kyoto 606-8502, Japan}

\email{tamura@kusastro.kyoto-u.ac.jp}
\begin{abstract}

Color gradients in elliptical galaxies in distant clusters
($z=0.37-0.56$) are examined by using the archival deep imaging data of
Wide Field Planetary Camera 2 (WFPC2) on-board the Hubble Space
Telescope (HST). Obtained color gradients are compared with the two
model gradients to examine the origin of the color gradients. In one
model, a color gradient is assumed to be caused by a metallicity
gradient of stellar populations, while in the other one, it is caused by
an age gradient. Both of these model color gradients reproduce the
average color gradient seen in nearby ellipticals, but predict
significantly different gradients at a redshift larger than $\sim$0.3.
Comparison between the observed gradients and the model gradients
reveals that the metallicity gradient is much more favorable as the
primary origin of color gradients in elliptical galaxies in clusters.
The same conclusion has been obtained for field ellipticals by using
those at the redshift from 0.1 to 1.0 in the Hubble Deep Field-North by
Tamura et al. (2000). Thus, it is also suggested that the primary origin
of the color gradients in elliptical galaxies does not depend on galaxy
environment.

\end{abstract}


\section{INTRODUCTION}

It has been known that nearby elliptical galaxies have color gradients;
colors in an elliptical galaxy gradually become bluer with increasing
radius (e.g., Vader et al. 1988; Franx, Illingworth, \& Heckman 1989;
Peletier et al. 1990a; Peletier, Valentijn, \& Jameson 1990b, Goudfrooij
et al. 1994; Michard 1999). Since many of elliptical galaxies show
radial gradients in metal absorption line strengths such as Mg$_{2}$,
Fe$_{1}$(5270 \AA) and Fe$_{2}$(5335 \AA) (e.g., Carollo, Danziger, \&
Buson 1993; Davies, Sadler, \& Peletier 1993; Gonzalez 1993; Kobayashi
\& Arimoto 1999), the origin of the color gradients has been naively
interpreted to be the metallicity gradients.

However, such an interpretation for the origin of the color gradient is
premature, because both metallicity gradient and age gradient in stellar
population can cause the same color gradient, and we cannot distinguish
the cause for the gradient.  This is called {\it age-metallicity
degeneracy}, which was originally pointed out by Worthey, Trager, \&
Faber (1996) in terms of the origin of the color-magnitude relation of
nearby elliptical galaxies (see also Arimoto 1996).
In order to break this degeneracy and to know the primary origin of the
color gradients in elliptical galaxies, comparing the observed color
gradients in distant ellipticals with predicted model gradients caused
by either the metallicity gradient or the age gradient is a very
effective approach, as was successful for examining the origin of the
color-magnitude (CM) relation (Kodama \& Arimoto, 1997).  Tamura et al.
(2000; hereafter called Paper I) constructed the two models both of
which reproduce a typical color gradient of elliptical galaxies at $z=0$
using a population synthesis model.  In one model, the mean metallicity
of the stellar population decreases with increasing radius at a fixed
old mean age. While in the other one, the mean age decreases with a
radius at a fixed mean metallicity.  These models were then made evolve
back in time.  The evolution of color gradients thus predicted are
confronted with the observed ones in distant ($z = 0.1 \sim 1.0$)
ellipticals sampled from the {\it Hubble Deep Field-North} (HDF-N;
Williams et al. 1996).  As a result, Paper I found that the metallicity
gradient is the primary origin of color gradients and the age gradient
model cannot reproduce the observed gradient at such redshift.

The elliptical galaxies in the HDF-N, however, are only those in field
environment. It has never been obvious that ellipticals in clusters
evolve similarly as those in field.  In rich clusters, it has been found
that the color-magnitude relation still holds even at around $z \sim 1$
(e.g., Stanford, Eisenhardt, \& Dickinson 1998) and these observational
results seem to favor the classical monolithic collapse scenario
associated with the galactic wind and high-$z$ formation (e.g., $z > 3$)
of elliptical galaxies (e.g., Kodama et al. 1998).  However, this kind
of evolution has not been established for ellipticals in lower density
environment (but see Kodama, Bower, \& Bell 1998).  Some predictions
either theoretically or observationally show that field ellipticals
formed by recent (at $z \leq 1$) merging processes (e.g., Baugh, Cole,
\& Frenk 1996; Barger et al. 1999).  An internal structure of a galaxy
such as a metallicity gradient and an age gradient must depend on its
formation process.  If cluster ellipticals pass different formation
histories from those for field ellipticals, their internal structures,
thus the origin of the color gradients, may not be the same. Or some
environmental effects on color gradients may exist. Thus, the same
approach is needed for cluster ellipticals to clarify the origin of
their color gradients.

It is noted that dust extinction in elliptical galaxies may also have
some effects on the color gradients (Goudfrooij \& de Jong 1995; Wise \&
Silva 1996; Silva \& Wise 1996).
However, about half of the detection towards ellipticals in far infrared
with IRAS are around $3\sigma$ threshold and confirmation is needed to
be definitive (Bregman et al. 1998). In addition, spatial distribution
of dust in an elliptical galaxy as well as dust mass which could affect
a color gradient are not established yet. These are still open problems
and will be examined in detail in our forthcoming papers.  Therefore, in
this paper, we have chosen to focus on age and metallicity effects only.

This paper is organized as follows. The sample selection and data
analysis of elliptical galaxies are described in \S~2. Histograms of
color gradients are presented in \S~3 together with the representative
color profiles of the sample ellipticals.  Discussion is given in \S~4.
The cosmological parameters adopted throughout this paper are the same
as those in Paper I; $H_{0} = 50$ km s$^{-1}$ Mpc$^{-1}$,
$\Omega_{0}=0.2$ and $\Lambda = 0$.

\section{DATA AND SAMPLE SELECTION}

In order to examine color gradients in elliptical galaxies in distant
clusters, deep imaging data with a high angular resolution are necessary
in more than two bands.  Thus, we choose to use the archival data taken
with the WFPC2 on-board the HST.  
Smail et al. (1997) obtained deep imaging data of 11 distant rich
clusters at the redshift from 0.37 to 0.56, most of which have exposure
times more than 10,000 sec to examine their morphology in detail, and
presented catalogs of the detected objects. In this paper, their reduced
imaging data which are available at their web
site\footnote{http://star-www.dur.ac.uk/\~{}irs/morphs.html} are
used. From these data, we select six clusters whose images were taken in
two bands to obtain galaxy colors. Cluster IDs and their basic
properties taken from Smail et al. (1997) are given in Table 1.  Among
these clusters, Cl 0024$+$16, Cl 0016$+$16, and Cl 0054$-$27 are
classified as high concentration clusters and Cl 0939$+$47 and Cl
0412$-$65 are low concentration clusters (Dressler et al. 1997).  For
A370 and Cl 0939$+$47, the images of their outer fields were taken.
Therefore, environmental effects on color gradients may be examined.

Early type galaxies (E, E/S0, S0/E) in these clusters are sampled based
on the catalog by Smail et al. (1997). Since our main interest is to
examine their color gradients, high signal-to-noise ratio (S/N) is
required and thus galaxies brighter than the apparent magnitude of 21
mag in $I_{814}$ band are selected in all of the sample clusters. This
apparent magnitude roughly corresponds to the absolute magnitude of
$-20$ mag in $V$ band. Our sample galaxies are listed in Table 2 with
their basic parameters, most of which are taken from Smail et al.
(1997). In each cluster, the sample galaxies form the tight CM relations
as shown in Figure 1, though the relations in A370, Cl 0939$+$47, and Cl
0412$-$65 are somewhat loose. Since Cl 0939$+$47 and Cl 0412$-$65 are
classified as low concentration clusters, and the outer regions of the
clusters were imaged for A370 and Cl 0939$+$47, some environmental
effects on the CM relations might be seen. However, a discussion for
this subject is beyond our scope in this paper. In the figure, a solid
square with an open circle shows the object removed from the following
analysis, because their color gradients could not be derived accurately
enough to use our discussion due to their low signal-to-noise ratios
(S/Ns) or due to their close positions to an edge of the image.

\section{PROFILES AND COLOR GRADIENTS}

In deriving color profiles of the sample galaxies, we should take care
of a positional coincidence between a blue image and a red image, and a
difference of the sizes of point spread functions (PSFs) between the two
images. (One pixel corresponds to $ 0^{\prime\prime}_{\cdot}10$ for the
Wide Field Cameras and $ 0^{\prime\prime}_{\cdot}046$ for the Planetary
Camera.) Positional displacement between a blue image and a red one
causes spurious asymmetry of a color distribution in a galaxy and has a
serious effect on the color gradient. We examined the centroids of the
sample galaxies in both blue and red images and corrected a
displacement, if a systematic shift between the two images in each
cluster was found.  The sizes of PSFs should be also estimated and be
adjusted between blue and red images.  We simulated the PSF at each
position of a sample galaxy in each band by using TinyTim v4.4 (Krist
1995; Krist \& Hook 1997), and corrected the difference of the sizes
between the two images, which were estimated by fitting a Gaussian to
the PSFs. Sky value was determined by ``mode'' in an annulus with an
inner radius of $9^{\prime\prime}$ and a width of 3$^{\prime\prime}$ by
using the ``phot'' task in the IRAF apphot package and the obtained sky
was subtracted.

We next made azimuthally averaged radial surface brightness profiles in
both blue and red images with a radial sampling of 0.$^{\prime\prime}$1. 
These profiles in each galaxy are made along the ellipses fitted with a
fixed position angle to the isophotes of the galaxy images in the red
band. (The position angles are taken from Table 2A by Smail et al.
(1997).) Thus the surface brightness profiles in the two bands are
derived based on the same fitted elliptical isophotes.
In Figure 2, representative surface brightness profiles of our sample
galaxies are shown.  The profiles of the brightest, intermediate, and
the faintest sample galaxies in each cluster are shown in top, middle,
and bottom panels, respectively.  The profiles of almost all the
galaxies in our sample are well represented by the $r^{1/4}$ law. To
derive color profiles, the surface brightness profile in the red band is
subtracted from the profile in the blue band.  Figure 3 shows the
resulting color profiles with the same arrangement as in Figure 2.  An
error bar to each data point includes a photometric error, a local sky
subtraction error (1$\%$ of a sky value around each object is adopted),
and a dispersion of colors along each elliptical isophote. It is
important to describe here the two dimensional color distribution in the
sample galaxies.  The color maps were constructed by dividing the blue
image by the red image after adjusting the positional displacement and
difference of the sizes of the PSFs. Almost all the color maps show
smooth color distribution and do not show any asymmetric or peculiar
features. Thus the color profiles well represent the color distribution
in each galaxy.

Finally, slopes of the color profiles, i.e., color gradients, in the
sample galaxies are derived by the least square fitting within effective
radii, which are obtained from the surface brightness profiles in the
red band assuming the $r^{1/4}$ law. The fitting for the color gradients
are done after removing data points with errors larger than 0.3 mag. We
do not derive the gradients of the profiles with accepted data points
fewer than 3. The average number of the data points used for the fitting
is about 8 and the rejected objects are minor.
Resulting color gradients in the sample galaxies are listed in Table 2
with the 1$\sigma$ fitting errors. Figure 3 indicates resulting slopes
of the color gradients as well as the slopes with $\pm 1 \sigma$. In
this figure, abscissa refers to logarithm of a radius normalized by the
outermost radius ($r_{f}$) used for the fitting in each color profile.
For most of the sample galaxies, $r_{f}$ is roughly equal to $r_{e}$.

Figure 4 shows a histogram for the color gradients of the sample
galaxies in each cluster. Each bin of the histograms is set to be 0.2
mag/dex which is comparable to the average value of the fitting error. 
It is found that the distributions of the gradients are very narrow
except for a few outliers, which are \#535 and \#738 in Cl 0024$+$16,
\#2005 in Cl 0939$+$47, and \#2050 in Cl 0016$+$16. The former three are
significantly out of the CM relation towards the blue side, but the last
one is almost on the relation. Considering that the range of the
distribution of the color gradients is comparable with or only slightly
larger than the estimated error for the slopes, the intrinsic
dispersions of the color gradients must be considerably small. (The
dispersion of the color gradients in nearby elliptical galaxies is about
0.04 mag/dex (Peletier et al. 1990a; Paper I).) It is intriguing that
the color gradients of elliptical galaxies are uniform even at
intermediate redshift. Furthermore, this encourages the comparison
between model gradients and observed gradients in distant clusters,
despite rather large errors for the observed slopes.

\section{ORIGIN OF COLOR GRADIENTS IN CLUSTER ELLIPTICALS}

\subsection{Models}

In order to examine whether the origin of the color gradient is the
stellar metallicity or the age, we adopt the same approach as that in
Paper I and the reader should refer to it in detail.  We briefly
summarize it here.

An observed color gradient can be reproduced by either a metallicity
gradient or an age gradient of stellar populations in an elliptical
galaxy at $z=0$.  However, since the color gradient caused by a
metallicity gradient is expected to follow a different evolution from
that by an age gradient, the origin of the color gradients can be found
by comparing the observed ones at high redshift with those predicted by
model. For this purpose, using the population synthesis model (Kodama \&
Arimoto 1997; Kobayashi, Tsujimoto, \& Nomoto 2000), we construct the
two model galaxies; one model galaxy has the color gradient made by the
pure metallicity gradient (thereafter called {\it metallicity gradient
model}) without age gradient, and the other made by the pure age
gradient ({\it age gradient model}) without metallicity gradient. In the
metallicity gradient model, the metallicity gradient is produced by
assuming that a galactic wind blowed later in the inner region in an
elliptical galaxy; star formation continued longer and thus the mean
stellar metallicity became higher at the inner region. For the age
gradient model, star formation started earlier in the inner region and
thus the mean age of stellar populations is older than that in the outer
region. The stellar population in each region in an elliptical galaxy is
assumed to be made by a single burst and to evolve independently of
other regions. Model parameters used here are set to the same as those
in Paper I, which are chosen so as to reproduce the typical color
gradient at $z=0$. The mean value of $\Delta(B-R)/\Delta \log r$ of
$-0.09$ mag$/$dex obtained by Peletier et al. (1990a) is adopted as the
typical color gradient at $z=0$. Note that these model galaxies must be
old (8 $\sim$ 15 Gyr) to reproduce colors in nearby elliptical galaxies. 
Then, we calculate the spectral evolution in each region of the model
galaxies and their color gradients at any redshifts using the response
functions including each filter on the HST.
It should be emphasized that we do not intend to study physical
formation process of elliptical galaxies in this paper, but aim at
depicting the evolution of the color gradient caused by either
metallicity gradient or age gradient to be compared with the observed
ones. Actual physical process that made the metallicity/age gradient may
be different from our brief ideas presented in the model description.
However it is not a problem here, because once such gradient formed,
subsequent evolution of the stellar population is unique and does not
depend on the formation process.

The two lines in each panel of Figure 5 show the evolutionary track of
the model color gradients; the solid curve indicates the evolution for
the metallicity gradient model and the dotted curve for the age gradient
model. The model color gradient by the metallicity gradient is almost
constant with a redshift within $z \sim 1$, while that by the age
gradient changes abruptly and shows a quite steep gradient even at
$z=0.3$. We will compare the model gradients with the observed ones in
the next subsection.

\subsection{Model vs observation}

The mean values of the color gradients in each cluster sample are
plotted at their redshifts in Figure 5. An error bar attached to each
point indicates a mean error of the gradients in each cluster. As
clearly shown, the metallicity gradient is much more favorable as the
origin of the color gradients. This result does not depend on
cosmological parameters or parameters for an evolutionary model of
galaxy within a reasonable range, and does not change even if we
consider the dispersion of the color gradients in the sample galaxies
(see Figure 4) and that in nearby ellipticals ($\sim 0.04$ mag/dex).
Although the sample galaxies of which memberships in the clusters are
spectroscopically confirmed are minor (Dressler et al. 1999), background
or foreground contaminations are not expected to affect the result for
the origin of the color gradients, because the result does not change
even if we remove the galaxies which significantly deviate from the CM
relation in each cluster. The color gradients in several sample galaxies
may be affected by other galaxies close to them, and the color profile
of a galaxy which locates close to an edge of the chip or on a joint
between the cameras may be somewhat spurious. However, our result still
holds even after removing the galaxies which may suffer from these
effects. 

Considering the result in Paper I, in both cluster and field, the
primary origin of the color gradients in elliptical galaxies is
considered to be the stellar metallicity. However, it is interesting to
point out that the mean values of the color gradients seem to deviate
upwards from the line for the metallicity gradient model.
Our models are calibrated by the color gradients seen in nearby
ellipticals by Peletier et al. (1990a), in which most of the sample
ellipticals reside in field or group environment. Therefore, the upward
deviation might indicate an environmental effect on the color gradients
of elliptical galaxies between in rich clusters and in field. However,
the correlation between the mean value and the degree of the
concentration in each cluster is not seen.  In addition, the mean
gradients of the clusters of which outer field images were taken do not
show larger values than others.  Further detailed study on the color
gradients in cluster ellipticals and field ones at $z=0$ as well as at
high redshift should be done in the future.


\acknowledgements

We would like to thank C. Kobayashi, N. Arimoto, and T. Kodama for
fruitful collaboration in Paper I. This work was financially supported
in part by Grant-in-Aid for the Scientific Research (No. 11740123) by
the Ministry of Education, Science, Sports and Culture of Japan.

\begin{table}
\tablenum{1}

 \caption{Sample Clusters}

 \vspace{5mm}

 \begin{tabular}{cccccc} \hline\hline
   ID & $z$ & \multicolumn{2}{c}{Image Center} & Exposure (sec) & 
   Exposure(sec) \\ 
      & & $\alpha$(J2000) & $\delta$(J2000) & (F555W)\tablenotemark{a} 
      & (F814W) \\ \hline
   A370 Field 2 & 0.37 & 02h40m01.1s 
   & $-01^{\circ}36^{\prime}45^{\prime\prime}$ & 8000 & 12600 \\
   Cl 0024$+$16 & 0.39 & 00h26m35.6s &
   $+17^{\circ}09^{\prime}43^{\prime\prime}$ & 23400 & 13200 \\
   Cl 0939$+$47 Field 2 & 0.41 & 09h43m02.5s 
   & $+46^{\circ}56^{\prime}07^{\prime\prime}$  & 4000 & 6300 \\
   Cl 0412$-$65 & 0.51 & 04h12m51.7s &
   $-65^{\circ}50^{\prime}17^{\prime\prime}$ & 12600 & 14700 \\
   Cl 0016$+$16 & 0.55 & 00h18m33.6s &
   $+16^{\circ}25^{\prime}46^{\prime\prime}$ & 12600 & 16800 \\
   Cl 0054$-$27 & 0.56 & 00h56m54.6s &
   $-27^{\circ}40^{\prime}31^{\prime\prime}$ & 12600 & 16800 \\ \hline\hline
 \end{tabular}
 \tablenotetext{a}{For Cl 0024$+$16, the exposure time in the F450W band
 image.}

\end{table}

\clearpage

\begin{table}
\tablenum{2}
\caption{Sample galaxies}

\begin{itemize}
 \item A370
\end{itemize}

\vspace{2mm}

 \begin{tabular}{ccccc} \hline\hline
   ID & $I_{814}$ & $V_{555} - I_{814}$ &
   $\Delta(V_{555}-I_{814})/\Delta$log$r$ & N\tablenotemark{a} \\ 
      & (mag) & (mag) & (mag$/$dex) & \\ \hline
   192 & 18.712 & 1.897 &    0.11$\pm$0.06 & 16  \\
   221 & 20.828 & 2.345 &    0.11$\pm$0.39 &  4  \\
   230 & 18.847 & 1.578 & $-$0.02$\pm$0.12 &  7  \\
   231 & 19.641 & 1.926 &    0.11$\pm$0.14 &  6  \\
   232 & 18.911 & 2.067 & $-$0.08$\pm$0.12 &  8  \\
   265 & 20.491 & 1.786 & $-$0.09$\pm$0.20 &  6  \\
   289 & 20.539 & 1.575 &    0.17$\pm$0.16 &  5  \\
   351 & 20.562 & 1.602 & $-$0.08$\pm$0.11 &  9  \\
   377 & 18.977 & 2.021 & $-$0.04$\pm$0.08 & 17  \\
   458 & 20.299 & 1.817 &    0.11$\pm$0.29 &  6  \\
   469 & 20.346 & 0.915 &      $-$         & $-$ \\
   487 & 19.071 & 1.867 &    0.00$\pm$0.08 & 16  \\
  2024 & 20.955 & 2.343 & $-$0.21$\pm$0.40 &  4  \\ \hline\hline

 \end{tabular}
 \tablenotetext{a}{Number of the data points in a color profile for
  deriving a color gradient.}
\end{table}

\begin{table}
\tablenum{2}
\caption{---Continue}

\begin{itemize}
 \item Cl 0024$+$16
\end{itemize}

\vspace{2mm}

 \begin{tabular}{ccccc} \hline\hline
   ID & $I_{814}$ & $B_{450} - I_{814}$ &
   $\Delta(B_{450}-I_{814})/\Delta$log$r$ & N \\ 
      & (mag) & (mag) & (mag$/$dex) & \\ \hline

   89 & 19.947 & 3.199 & $-$              & $-$ \\
  112 & 19.519 & 3.278 & $-$0.00$\pm$0.11 & 13 \\
  113 & 18.892 & 3.404 & $-$0.13$\pm$0.09 & 14 \\
  137 & 20.827 & 3.391 & $-$0.52$\pm$0.45 & 4 \\
  145 & 20.717 & 3.251 &    0.18$\pm$0.38 & 4 \\
  147 & 19.411 & 3.290 & $-$0.27$\pm$0.12 & 10 \\
  169 & 20.795 & 3.255 & $-$              & $-$ \\
  179 & 19.654 & 3.559 & $-$0.01$\pm$0.18 & 7 \\
  261 & 19.623 & 3.310 &    0.06$\pm$0.23 & 7 \\
  268 & 19.400 & 3.296 &    0.14$\pm$0.19 & 8 \\
  280 & 18.200 & 3.481 & $-$0.22$\pm$0.06 & 30 \\
  294 & 19.959 & 3.296 &    0.42$\pm$0.21 & 6 \\
  304 & 18.469 & 3.419 & $-$0.28$\pm$0.06 & 26 \\
  327 & 20.759 & 3.173 & $-$              & $-$ \\
  334 & 19.567 & 3.389 & $-$0.03$\pm$0.15 & 8 \\
  337 & 20.069 & 3.348 &    0.16$\pm$0.37 & 4 \\
  342 & 18.680 & 3.412 & $-$0.05$\pm$0.08 & 20 \\
  343 & 18.348 & 3.506 & $-$0.28$\pm$0.05 & 30 \\ 
  353 & 20.557 & 3.220 &    0.01$\pm$0.40 & 4 \\
  362 & 18.901 & 3.367 & $-$0.04$\pm$0.12 & 8 \\
  365 & 18.309 & 3.403 & $-$0.17$\pm$0.06 & 27 \\ \hline\hline

 \end{tabular}
\end{table}

\begin{table}
\tablenum{2}
\caption{--- Continue}

\begin{itemize}
 \item Cl 0024$+$16 --- Continue
\end{itemize}

\vspace{2mm}

 \begin{tabular}{ccccc} \hline\hline
   ID & $I_{814}$ & $B_{450} - I_{814}$ &
   $\Delta(B_{450}-I_{814})/\Delta$log$r$ & N \\ 
      & (mag) & (mag) & (mag$/$dex) & \\ \hline

  403 & 19.349 & 3.317 & $-$0.55$\pm$0.14 & 11 \\
  419 & 19.573 & 3.444 & $-$0.13$\pm$0.19 & 7 \\
  479 & 20.768 & 3.170 &    0.14$\pm$0.39 & 4 \\ 
  514 & 19.883 & 3.325 & $-$0.34$\pm$0.19 & 7 \\
  521 & 20.259 & 3.318 &    0.22$\pm$0.29 & 5 \\
  535 & 20.343 & 2.501 & $-$1.68$\pm$0.12 & 17 \\
  573 & 18.353 & 3.424 & $-$0.19$\pm$0.20 & 10 \\
  590 & 20.201 & 3.385 & $-$              & $-$ \\
  621 & 18.660 & 3.405 & $-$0.20$\pm$0.07 & 16 \\
  653 & 19.090 & 3.311 & $-$0.44$\pm$0.12 & 10 \\
  669 & 20.130 & 3.282 & $-$0.38$\pm$0.24 & 6 \\
  675 & 20.709 & 3.216 & $-$0.07$\pm$0.37 & 4 \\
  678 & 20.298 & 3.256 & $-$0.15$\pm$0.30 & 5 \\
  685 & 20.683 & 3.196 & $-$0.30$\pm$0.38 & 4 \\
  738 & 20.522 & 3.208 & $-$3.23$\pm$0.35 & 10 \\
  796 & 19.109 & 3.450 & $-$0.22$\pm$0.11 & 13 \\
  876 & 19.669 & 3.315 & $-$0.15$\pm$0.23 & 5 \\
  934 & 20.382 & 2.259 & $-$              & $-$ \\
 3006 & 20.944 & 3.118 &    0.13$\pm$0.14 & 18 \\
 3012 & 20.475 & 3.367 & $-$0.12$\pm$0.15 & 12 \\ \hline\hline

 \end{tabular}
\end{table}

\begin{table}
\tablenum{2}
\caption{---Continue}

\begin{itemize}
 \item Cl 0939$+$47
\end{itemize}

\vspace{2mm}

 \begin{tabular}{ccccc} \hline\hline
   ID & $I_{814}$ & $B_{450} - I_{814}$ &
   $\Delta(B_{450}-I_{814})/\Delta$log$r$ & N \\ 
      & (mag) & (mag) & (mag$/$dex) & \\ \hline

   31 & 20.452 & 2.040 & $-$              & $-$ \\
   53 & 20.247 & 2.770 & $-$0.17$\pm$0.31 & 5 \\
   86 & 18.769 & 1.822 &    0.01$\pm$0.04 & 11 \\
  270 & 19.993 & 1.866 & $-$0.18$\pm$0.12 & 10 \\
  337 & 20.507 & 2.209 &    0.02$\pm$0.20 & 6 \\
  404 & 19.670 & 2.344 & $-$0.01$\pm$0.19 & 7 \\
  426 & 20.014 & 1.947 & $-$0.10$\pm$0.15 & 7 \\
  429 & 20.372 & 2.211 &    0.18$\pm$0.31 & 5 \\
  512 & 20.972 & 1.817 &    0.04$\pm$0.24 & 6 \\
  515 & 20.208 & 1.997 & $-$0.04$\pm$0.13 & 9 \\
  566 & 20.557 & 2.010 & $-$              & $-$ \\
 2005 & 20.968 & 0.700 & $-$0.76$\pm$0.31 & 11 \\ \hline\hline

 \end{tabular}
\end{table}

\begin{table}
\tablenum{2}
\caption{---Continue}

\begin{itemize}
 \item Cl 0412$-$65
\end{itemize}

\vspace{2mm}

 \begin{tabular}{ccccc} \hline\hline
   ID & $I_{814}$ & $B_{450} - I_{814}$ &
   $\Delta(B_{450}-I_{814})/\Delta$log$r$ & N \\ 
      & (mag) & (mag) & (mag$/$dex) & \\ \hline

  431 & 20.835 & 2.715 &    0.03$\pm$0.22 & 7 \\
  432 & 18.992 & 2.044 & $-$0.08$\pm$0.07 & 18 \\
  471 & 20.074 & 2.254 &    0.29$\pm$0.16 & 4 \\
  472 & 19.571 & 2.330 &    0.11$\pm$0.10 & 9 \\
  635 & 19.612 & 2.300 &    0.06$\pm$0.10 & 14 \\
  657 & 19.597 & 2.265 &    0.06$\pm$0.10 & 7 \\
  682 & 20.960 & 2.183 &    0.11$\pm$0.16 & 7 \\
  695 & 20.440 & 2.213 & $-$0.04$\pm$0.16 & 4 \\
  772 & 20.489 & 0.839 &    0.23$\pm$0.06 & 14 \\ \hline\hline

 \end{tabular}
\end{table}

\begin{table}
\tablenum{2}
\caption{---Continue}

\begin{itemize}
 \item Cl 0016$+$16
\end{itemize}

\vspace{2mm}

 \begin{tabular}{ccccc} \hline\hline
   ID & $I_{814}$ & $B_{450} - I_{814}$ &
   $\Delta(B_{450}-I_{814})/\Delta$log$r$ & N \\ 
      & (mag) & (mag) & (mag$/$dex) & \\ \hline

  271 & 20.913 & 2.387 &    0.29$\pm$0.23 & 4 \\
  438 & 19.754 & 2.460 & $-$0.32$\pm$0.07 & 19 \\
  461 & 20.296 & 2.461 &    0.19$\pm$0.23 & 4 \\
  531 & 20.692 & 2.443 & $-$              & $-$ \\
  602 & 20.926 & 2.469 & $-$              & $-$ \\
  606 & 20.769 & 2.305 & $-$              & $-$ \\
  611 & 20.232 & 2.466 &    0.19$\pm$0.16 & 8 \\
  612 & 19.648 & 2.574 & $-$0.12$\pm$0.10 & 10 \\
  650 & 19.464 & 2.482 & $-$0.05$\pm$0.07 & 17 \\
  653 & 19.837 & 2.384 & $-$0.00$\pm$0.13 & 8 \\
  659 & 19.950 & 2.441 &    0.04$\pm$0.13 & 10 \\
  724 & 19.075 & 2.582 & $-$0.09$\pm$0.06 & 20 \\
  725 & 19.117 & 2.531 & $-$0.02$\pm$0.05 & 28 \\
  726 & 20.826 & 2.317 & $-$              & $-$ \\
  732 & 20.009 & 2.398 &    0.26$\pm$0.18 & 7 \\
  745 & 20.342 & 2.519 &    0.12$\pm$0.14 & 9 \\
  802 & 20.898 & 2.411 &    0.06$\pm$0.18 & 7 \\
  822 & 20.956 & 2.320 &    0.07$\pm$0.17 & 9 \\
  823 & 20.346 & 2.387 & $-$0.06$\pm$0.12 & 10 \\
  843 & 20.270 & 2.372 &    0.08$\pm$0.12 & 7 \\
  903 & 20.927 & 2.300 &    0.14$\pm$0.27 & 4 \\
 2026 & 20.643 & 2.187 & $-$0.32$\pm$0.19 & 5 \\
 2050 & 20.894 & 2.366 & $-$1.92$\pm$0.42 & 4 \\
 3002 & 18.894 & 2.122 & $-$              & $-$ \\ \hline\hline

 \end{tabular}
\end{table}

\begin{table}
\tablenum{2}
\caption{---Continue}

\begin{itemize}
 \item Cl 0054$-$27
\end{itemize}

\vspace{2mm}

 \begin{tabular}{ccccc} \hline\hline
   ID & $I_{814}$ & $B_{450} - I_{814}$ &
   $\Delta(B_{450}-I_{814})/\Delta$log$r$ & N \\
      & (mag) & (mag) & (mag$/$dex) & \\ \hline

  165 & 20.802 & 2.329 &    0.25$\pm$0.27 & 5 \\
  191 & 20.225 & 2.459 & $-$0.11$\pm$0.15 & 9 \\
  216 & 19.627 & 2.600 & $-$0.01$\pm$0.12 & 9 \\
  229 & 20.393 & 2.565 &    0.07$\pm$0.20 & 5 \\
  356 & 20.080 & 2.469 & $-$0.12$\pm$0.17 & 6 \\
  365 & 20.407 & 2.345 &    0.06$\pm$0.17 & 7 \\
  440 & 19.316 & 2.514 & $-$0.12$\pm$0.11 & 13 \\
  529 & 20.021 & 2.415 & $-$0.26$\pm$0.12 & 10 \\
  711 & 18.037 & 1.403 & $-$0.14$\pm$0.06 & 20 \\ \hline\hline

 \end{tabular}
\end{table}

\clearpage

\centerline{\bf Figure Caption}

\noindent
{Figure. 1 --- Color-magnitude (CM) diagrams for the sample galaxies in
the clusters. A solid square with an open circle indicates the object
whose color gradient cannot be obtained due to low S/N or the close
position to an edge of the image. ``\#2'' in the cluster ID refers to
the outer field of the cluster. }

\noindent 
{Figure. 2 --- Azimuthally averaged radial surface brightness profiles
of representative galaxies in the sample are presented. The profiles of
the brightest, intermediate, and the faintest sample galaxies in each
cluster are shown in top, middle, and bottom panels, respectively. Solid
squares are the profile in a blue band and open circles in a red
band. An object ID is shown at the upper left in each panel.}

\noindent 
{Figure. 3 --- Color profiles and fitted slopes of the representative
galaxies in the sample are indicated. The galaxies presented and the
arrangement of the panels are the same as those in Figure 2. Among the
three solid lines in each panel, the middle one shows the best fit
slope. Other two lines show the slopes with $\pm 1\sigma$ of the best fit. 
Abscissa refers to logarithm of a radius normalized by the
outermost radius ($r_{f}$) used for the fitting in each color profile.}

\noindent 
{Figure. 4 --- Histograms of the color gradients in the sample galaxies
in each sample cluster. A total number of the objects in each histogram
is shown at the upper left in each panel and a number in parentheses
indicates a total number of the sample galaxies in each cluster. Cluster
ID and its redshift are shown on the top of each panel.}

\noindent
{Figure. 5 --- A mean color gradient in each cluster versus redshift is
shown. A solid curve in each panel represents the evolutionary track of
the color gradient caused by the metallicity gradient and a dotted curve
shows the track by the age gradient (see text in detail).}

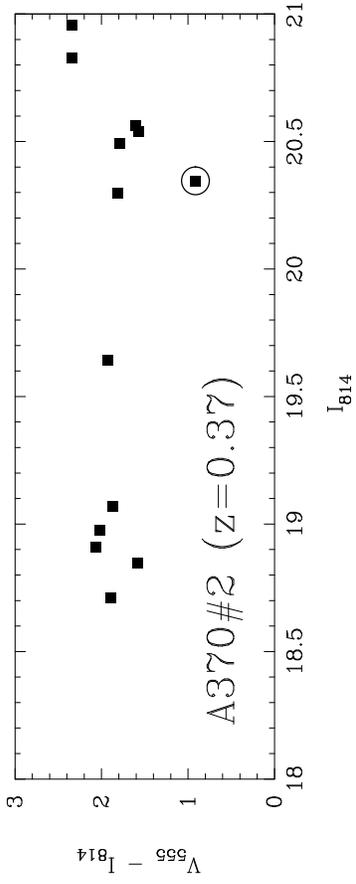
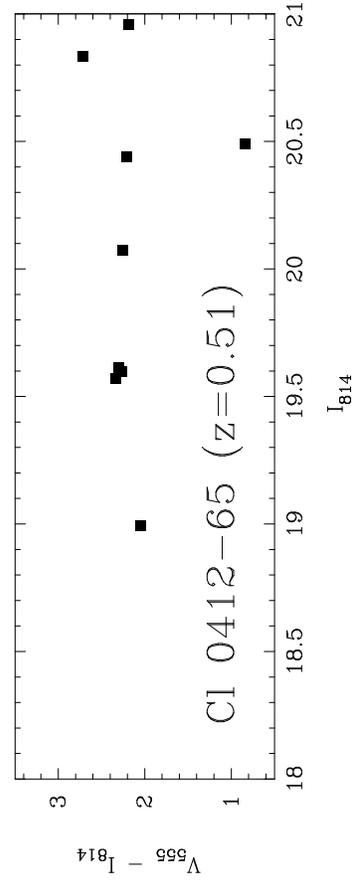
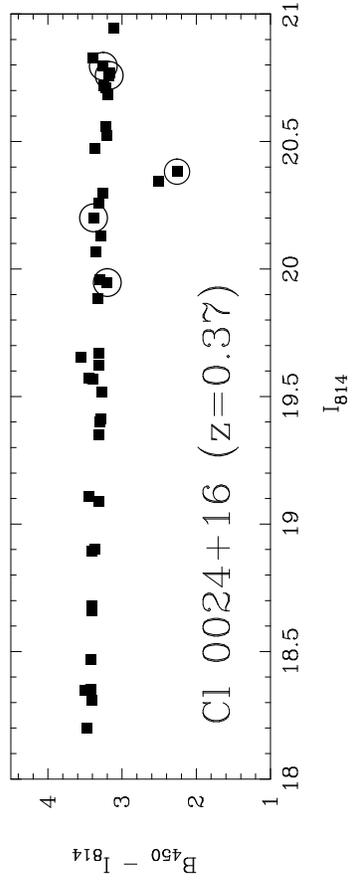
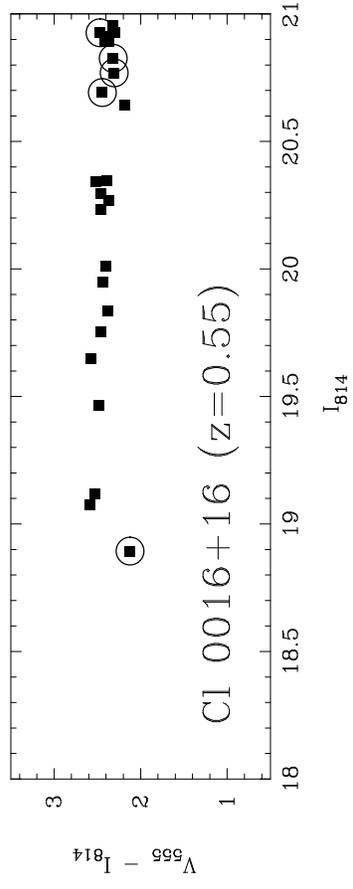
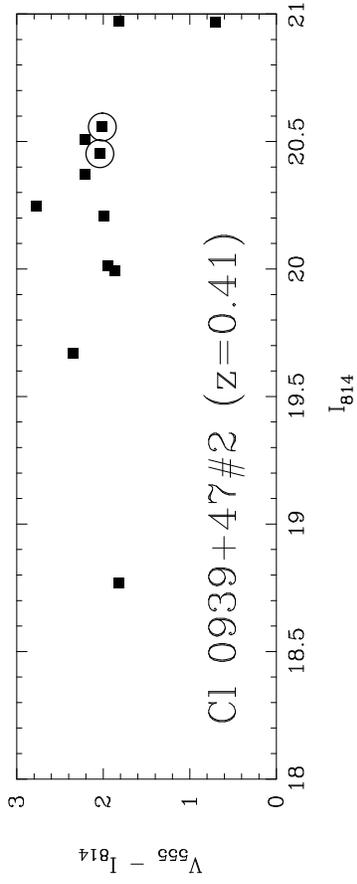
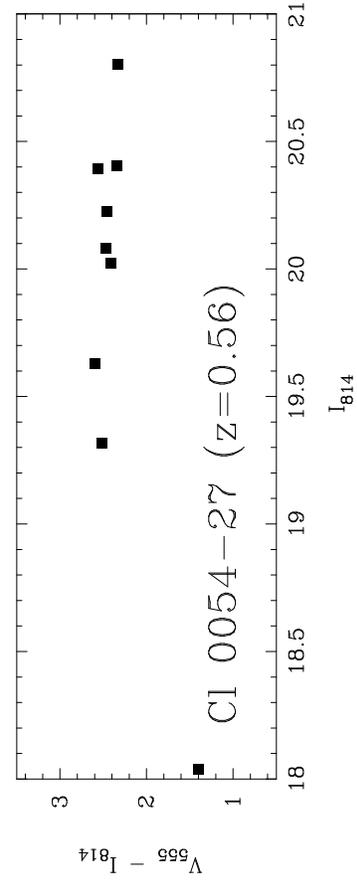
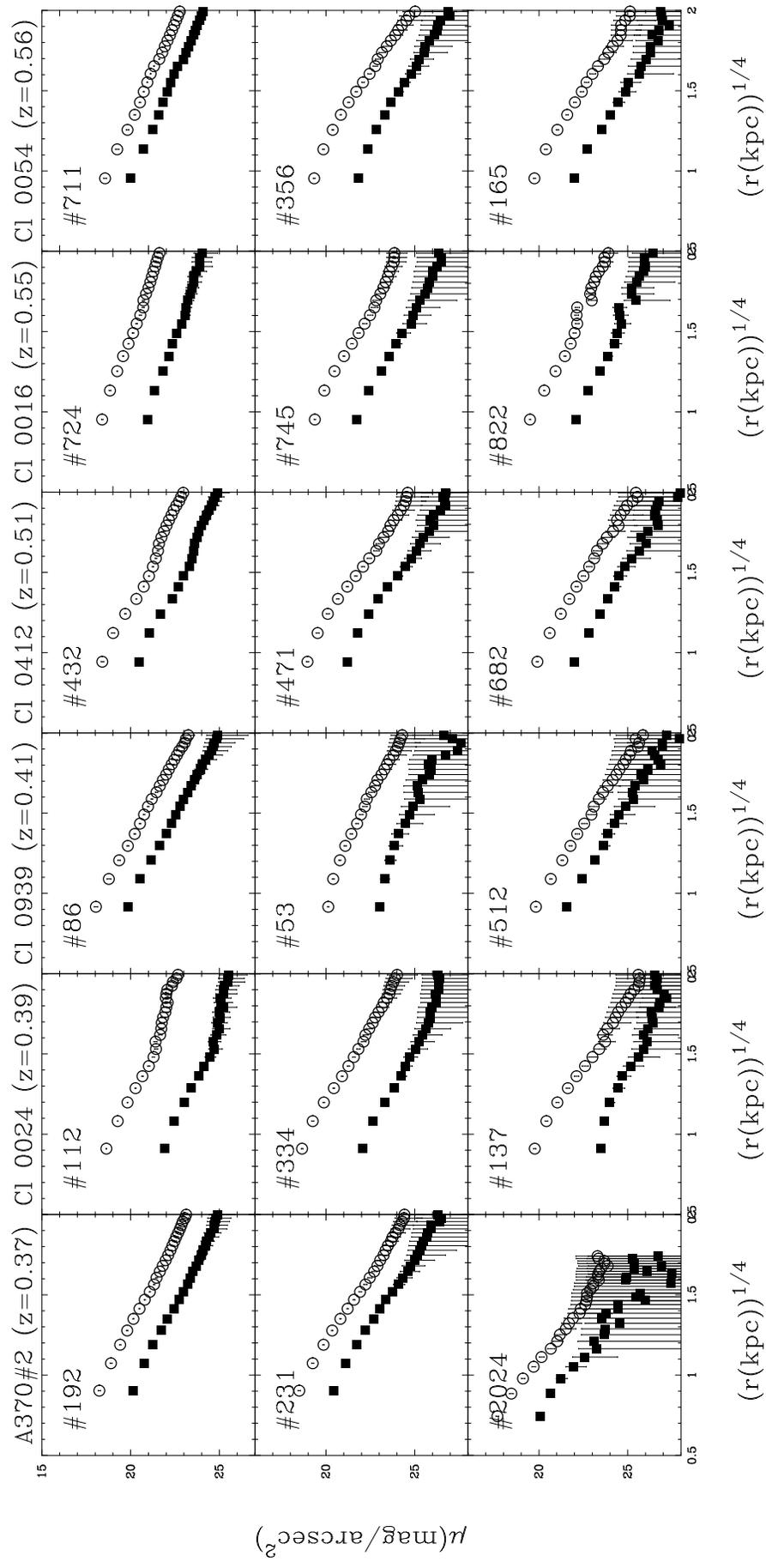
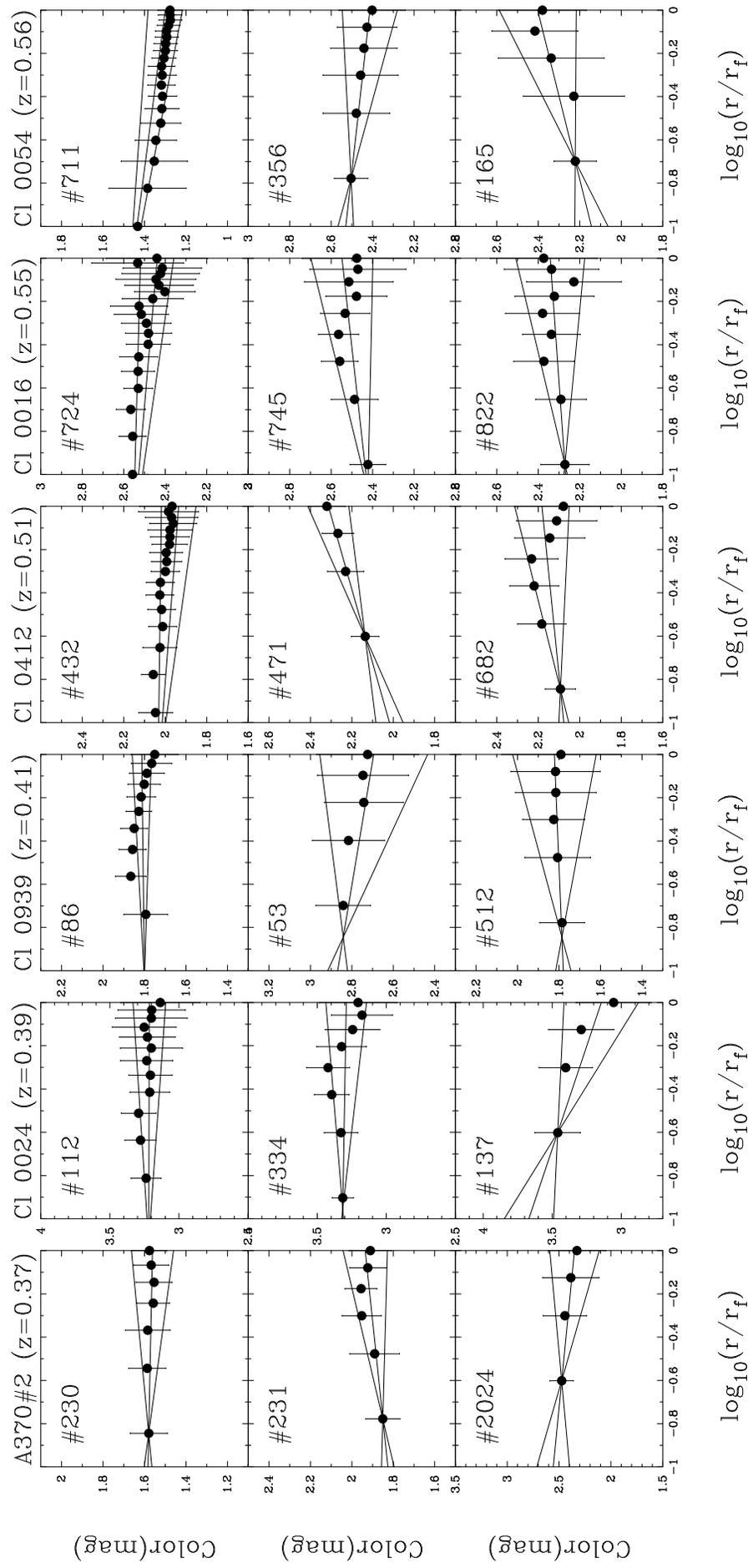
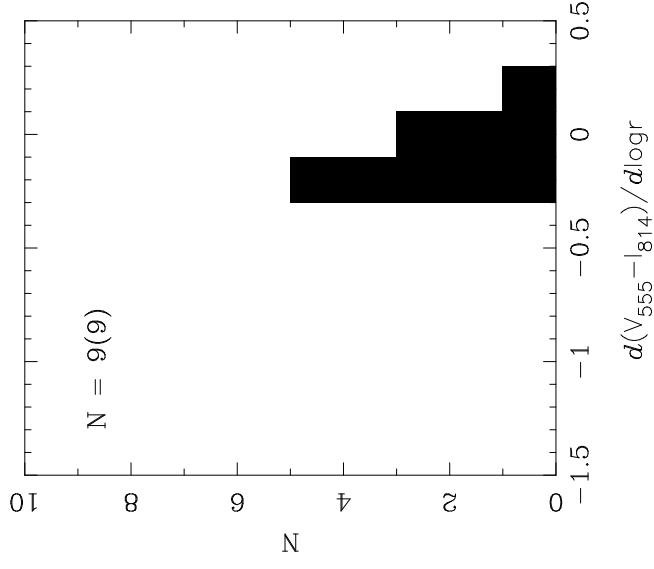
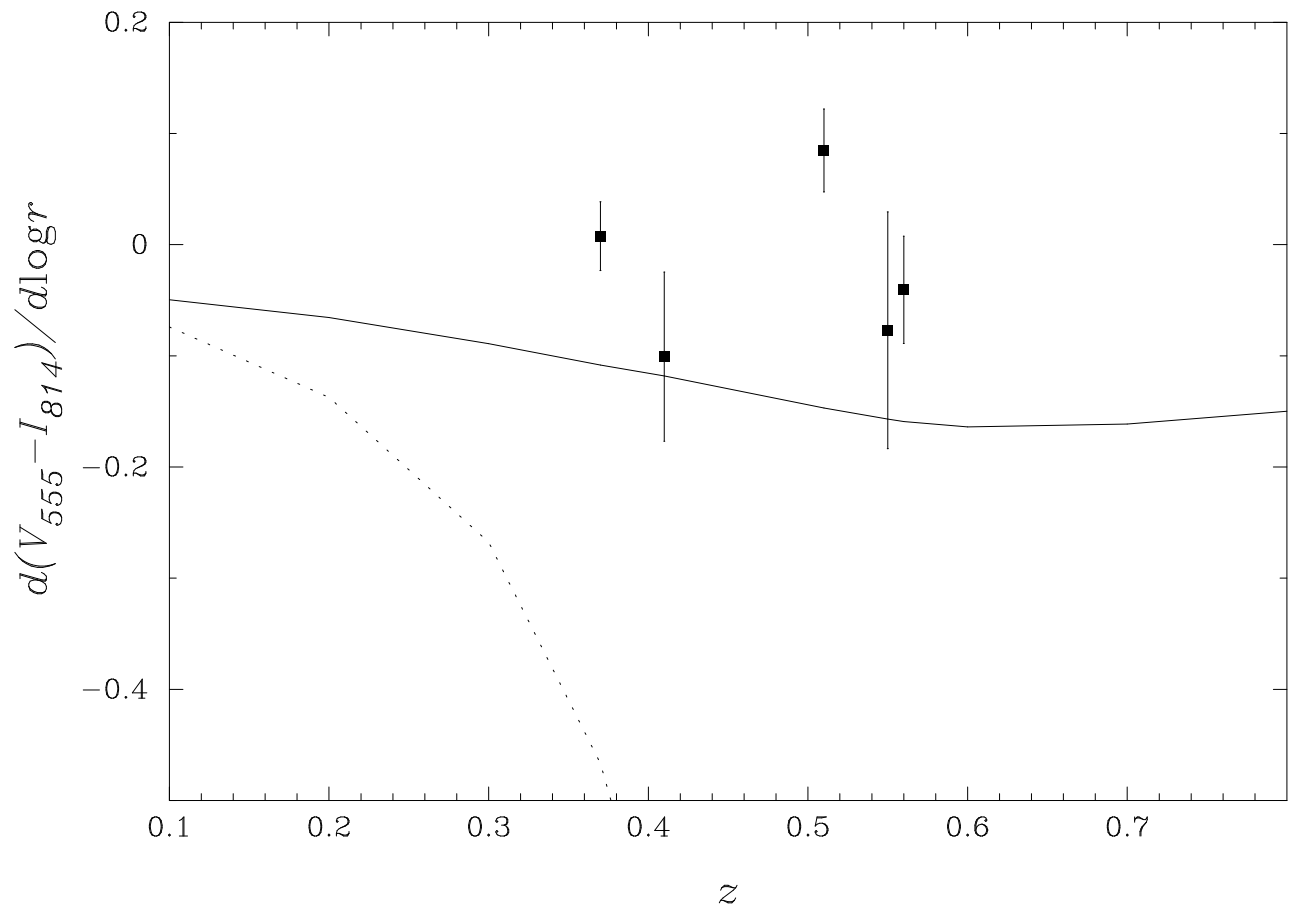
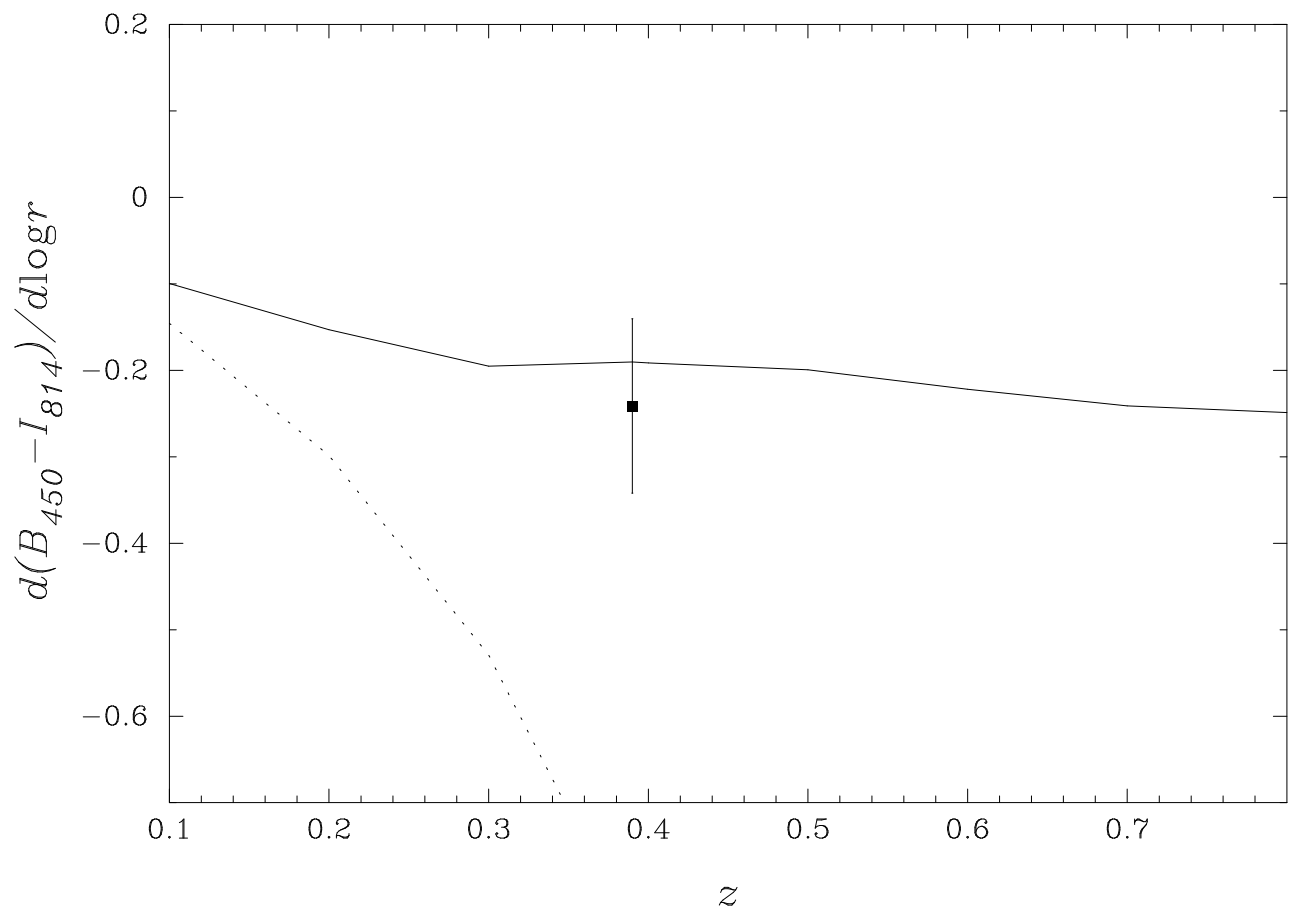
\end{document}